\documentclass{article}
\usepackage{amsmath,bbm,amssymb,color,amsfonts}
\usepackage{graphicx}
\newcommand{\ds}{\displaystyle}
\newcommand{\beq}{\begin{equation}}
\newcommand{\eeq}{\end{equation}}
\newcommand{\beqa}{\begin{eqnarray*}}
\newcommand{\eeqa}{\end{eqnarray*}}
\newcommand{\beqaa}{\begin{eqnarray}}
\newcommand{\eeqaa}{\end{eqnarray}}
\begin{document}

\begin{center}
{\Large \bf Exact Nonclassical Symmetry Solutions of Lotka-Volterra
Type Population Systems }
 \medskip\\
{\bf Phillip  Broadbridge~$^{1}$, Roman Cherniha~$^{2}$  and Joanna
Goard~$^{3}$}
\medskip\\
$^{1}\,$School of Engineering and Mathematical Sciences and
Institute of Math. for Industry- Kyushu University, La Trobe
University,\\
Bundoora VIC 3086, Australia\\
    email\textup{\nocorr: \texttt{P.Broadbridge@latrobe.edu.au}}\\
  $^2\,$ Institute of Mathematics
    National Academy of Science of Ukraine
    3, Tereshchenkivs'ka Street
    01004  Kyiv, Ukraine\\
\texttt{r.m.cherniha@gmail.com}
      \\
  $^3\,$School of Mathematics and Applied Statistics, University of Wollongong NSW 2522,
  Australia\\
  \texttt{joanna@uow.edu.au}
\end{center}

\begin{abstract}
New classes of conditionally integrable systems of nonlinear reaction-diffusion equations are introduced. They are obtained by extending a well known nonclassical symmetry of a scalar partial differential equation to a vector equation. New exact solutions of nonlinear predator-prey systems, related to the diffusive Lotka-Volterra system, are constructed.  An infinite dimensional class of exact solutions is made available. Unlike in the standard Lotka-Volterra system, in the absence of predators, the prey population has a finite carrying capacity, as in the Fisher equation. \\
\end{abstract}
Keywords: Nonlinear reaction-diffusion systems, Lotka-Volterra, predator-prey, nonclassical symmetries, integrability.\\

\section{Introduction}
Two-  and multi-component systems of nonlinear reaction-diffusion equations have well known applications to mobile interacting reagents and cells in chemical kinetics, biological morphogenesis, some biomedical processes  and population ecology \cite{mur-1989, mur2003, okubo, wa-2015, ku-na-ei-16}. Understanding of the dynamical behaviour of such systems has been developed largely from stability theory of steady states and bifurcation theory, reduction to travelling waves and numerical simulations. There are very many works devoted to these topics (see, the books cited above and,  e.g., recent papers \cite{alha-qu-19, lam-20} and references therein).\\

On the other hand, a  relatively small number of papers is devoted to the search for exact solutions of the systems of nonlinear reaction-diffusion equations arising in the applications mentioned above. To the best of our knowledge, the main attention in this direction was paid to the diffusive Lotka-Volterra (DLV) system, for which several exact solutions in  explicit forms were constructed \cite{rod-mimura-2000,ch-du-04,ch-dav-2011, ch-dav2013,ch-dav2021,hung12, chen-hung12}, many of which are summarized in the book \cite{Cherniha1}. There are also a few studies devoted to finding exact solutions of direct generalizations of the DLV system  arising in real-world applications \cite{pliukhin-15, ch-dav2019, ch-dav2021,ch-dav2020}. All the known solutions of the DLV system and its generalizations can be divided into two classes. The first one consists of travelling plane waves which make up an important class of solutions that are obtainable via the straightforward reduction to systems of ordinary differential equations (ODE).  The second class consists of exact solutions  obtained from classical or nonclassical ($Q$-conditional)  symmetry reductions. It should be pointed out that all the exact solutions presented in the works cited above were found after restricting the dependent density variables to (1+1)-dimensional time-space domains.

For mathematical modelling some processes in biology and ecology,  the variable diffusivity should be used.
The porous Fisher equation is a typical example of a scalar field model in that case \cite{mur-1989}).
In such a case, the corresponding reaction-diffusion system  is more complicated  and the problem of constructing exact solutions
is highly non-trivial, especially if one considers a real-world model in (1+2)- or higher-dimensional time-space. Some examples
in (1+1)-dimensional time-space are summarised  in books \cite{Cherniha1,polyanin-2012}. In this paper, we  concentrate on the systems
 with  variable diffusivities in (1+2)-dimensional time-space. That is particularly appropriate for population densities
 of the many species whose range of movement in the vertical direction is relatively insignificant.\medskip\\

General reaction-diffusion systems with nonlinearity in  both reaction and diffusion, offer more general possibilities.
 One pathway to explore those is to examine if known nonclassical symmetries of scalar equations can be extended to coupled vector systems.
 The notions of conditional and nonclassical symmetries, originating most prominently in the works of  Bluman and Cole \cite{Bluman},
  Fushchych \cite{Fushchych,Fush-88}, Olver \cite{Olve87}, Winternitz \cite{Levi89} and their co-authors
  eventually opened up the possibility of new reductions and solutions to practical partial differential equations that could not
   be obtained by Lie's classical algorithm. Notably,   techniques like the method of differential constraints
   of Yanenko \cite{yanenko-84},   the direct reduction method of Clarkson and Kruskal \cite{Clarkson} and some others
    (see for details Chapter 5 of book \cite{ch-se-pl-book}), which are not  based on symmetries, also can help in  constructing
     new non-Lie  solutions for  nonlinear PDEs arising in real-world applications.

  In \cite{Goard}, we found a class of  scalar nonlinear reaction-diffusion equations in 2+1 dimensions with a simple nonclassical symmetry that enables reduction to a pair of separated linear equations.
\begin{equation}
\label{RD}
\frac{\partial\theta}{\partial t}=\nabla\cdot[D(\theta)\nabla\theta]+R(\theta).
\end{equation}
In terms of the Kirchhoff flux potential
\begin{eqnarray}
\label{Kirchhoff}
\mu &=& \int _{\theta_j}^\theta D(\theta) d\theta, \\
\frac {1}{D(\theta(\mu))}\frac{\partial\mu}{\partial t}&=&L\mu+R(\theta(\mu)),
\end{eqnarray}
where $L$ is the 2D Laplacian operator. A solution of the form
\begin{equation}
\mu=e^{At}F({\bf x});~~~LF({\bf x})+\kappa F=0,
\label{reduction}
\end{equation}
satisfying a Helmholtz equation  is compatible with (\ref{RD}) if and only if the nonlinear diffusivity and nonlinear reaction term are related by
\begin{equation}
\label{reln}
R(\theta)=\kappa\mu+\frac{A\mu}{D}.
\end{equation}
If $D(\theta)$ is known, then the compatible $R(\theta)$ follows by direct integration as in (\ref{Kirchhoff}). In many applications, it is more important to specify $R(\theta)$ after which (\ref{reln}) must be solved as a differential equation for $\mu(\theta)$; subsequently $D(\theta)=\mu'(\theta)$. In practice, exact pairs $(D_{(n)},R_{(n)})$ are obtained by a  small number of iterations of the converging contraction map
\begin{equation}
\label{recursion}
D_{(n+1)}=\frac{A\int D_{(n)} d\theta}{R-\kappa\int D_{(n)} d\theta};~~D_{(0)}=-A/\kappa ,
\end{equation}
after which $R_{(n+1)}$ is obtained from $D_{(n+1)}$ as in (\ref{reln}).\\
In practice, this device works in any number of spatial dimensions and the Laplacian $L$ may be generalised to any linear differential operator acting on smooth functions of ${\bf x}\in \mathbb R^n$. In \cite{Triadis} we produced the only known exact solutions of temperature with Arrhenius combustion and diffusion in two and three dimensions. In \cite{hajek} we produced most of the very few known exact solutions for a diffusing population with the Verhulst logistic growth term or for a diffusing new competitive gene through a diploid population with cubic Huxley growth that arises from the Mendelian diploid inheritance. In \cite{Daly}, using the Kirchhoff operator $L=\nabla^2-\alpha \frac{\partial}{\partial z}$ we solved for water content in  unsaturated soil subject to plant root uptake. In \cite{Gallage} we took L to be a fourth-order Cahn-Hilliard diffusion operator to solve for phase bands in a solid/liquid mixture. In \cite{hajek2} we took L to be a variable-coefficient diffusion operator to solve for calcium diffusion on the spherical surface of a fertilised egg.\\

Now we begin to extend this formalism to isotropic {\em coupled} reaction-diffusion equations. We are primarily interested in two coupled equations that have various applications such as coupled heat and mass transfer, population ecology and activator/inhibitor enzymes for embryo morphogenesis. However, the same approach applies to any number of coupled equations.\\

In Section 2, the nonclassical reduction method is developed for two coupled reaction-diffusion equations. Nonclassical reductions allow for a much more general class of reaction terms.\\
In Section 3, exact oscillatory-in-time  solutions with spatial dependence, are provided for a cross-diffusion pursuit model with reaction terms that have similar properties to those of the classic Lotka-Volterra predator-prey system. In this case, the flux potentials are additively separable. The original Lotka-Volterra system was a pair of coupled ordinary differential equations  \cite{lot-1920,vol-1926}, not allowing for spatial variability in population densities. The extension of modified Lotka-Volterra systems to a pair of partial differential equations has been well studied. However, exact solutions with non-trivial variation in both space (especially in 2D case) and time, have been elusive.\\
In Section 4, another system of the modified Lotka-Volterra class, with multi-variate diffusion coefficients, is solved in the case of monotonic time dependent populations. This model follows from flux potentials that are multiplicatively separable.\\
Finally, in the conclusion, the progress is recapped, unsolved problems are identified and future investigations are suggested.

\section{Two coupled reaction-diffusion equations.}

Let us consider the two-component system of  reaction-diffusion equations
\begin{equation}
\label{coupled}
\frac{\partial\theta_j}{\partial t}=\frac{\partial}{\partial x^m}\left[{D_j}^k(\theta)\frac{\partial\theta_k}{\partial x^m}\right]+R_j({\bf \theta});~~ j,k=1,2; m=1,...,N.
\end{equation}
Hereafter $\theta=(\theta_1,\theta_2)$ is an unknown vector function, ${D_j}^k$ and $R_j$ are given smooth functions, and repeated indices will be summed.
The flux density $ {\bf J}^p$ of each population labelled $p=1,2$ will be assumed to be the gradient of a potential function,
\begin{eqnarray}
{\bf J}^p&=&-\nabla \mu_p(\theta_1,\theta_2)\\
&=&-\frac{\partial \mu_p}{\partial \theta_q}\nabla\theta_q\\
&=&-D_p~^q(\theta)\nabla\theta_q.
\end{eqnarray}
The condition for $d\mu_p$ to be an exact differential is simply
$$\frac{\partial }{\partial \theta_k}\frac{\partial \mu_p}{\partial \theta_j}-\frac{\partial }{\partial \theta_j}\frac{\partial \mu_p}{\partial \theta_k}=0$$
which is equivalent to
\begin{equation}
\label{exact}
\frac{\partial}{\partial \theta_k}D_p~^j=\frac{\partial}{\partial \theta_j}D_p~^k.
\end{equation}
In terms of the flux potentials, the system of reaction-diffusion equations is
\begin{equation}
\label{mueq}
{(D^{-1})_q}^p\frac{\partial\mu_p}{\partial t}=\frac{\partial \theta_q}{\partial t}=\nabla^2\mu_q+R_q({\bf\theta}({\bf \mu})).
\end{equation}
Such a general system may be either parabolic or hyperbolic in character. The latter case occurs when the diffusion matrix has pure imaginary eigenvalues. The only known fully integrable example is
\begin{eqnarray}
\nonumber
\frac{\partial u}{\partial t}=-\frac{\partial^2 v}{\partial x^2}+s(u^2+v^2)v,\\
\frac{\partial v}{\partial t}=\frac{\partial^2 u}{\partial x^2}-s(u^2+v^2)u.
\end{eqnarray}
In terms of the complex wave function $\psi=u+iv$, this is equivalent to the nonlinear Schr\"odinger equation,
$$i\frac{\partial\psi}{\partial t}=-\frac{\partial^2\psi}{\partial x^2}+s|\psi|^2\psi.$$
Beyond the integrable 2-vector equation in one space dimension, there is a conditionally integrable vector equation with any number $N$ of independent spatial variables, for which an exact time-dependent solution can be constructed from any solution of the linear matrix Helmholtz equation in $N$-dimensional space.\\
Beginning with a single scalar equation, wherein all indices $p$ and $q$ in the above are 1, (\ref{reln}) is the relation between nonlinear reaction rate and nonlinear diffusivity that allows the reaction diffusion equation to have a nonclassical symmetry with invariant surface condition
$$\mu_t=A\mu .$$ A reduced relationship among invariants $\mu e^{-At}$ and $x^i$ then results in (\ref{reduction}).
It becomes apparent that this algebraic construction still applies when $\mu$ is a vector, R is a vector, A is a constant square matrix and $\kappa$ is extended to a constant square matrix $M$. $e^{At}$ is defined in the usual way as a Taylor series
$$e^{At}=I+\sum_{n=1}^\infty \frac{(tA)^n}{n !}$$ after which we can take matrix components. For example if $A$ is skew-symmetric then $e^{At}$ is orthogonal.  The system of coupled reaction-diffusion equations is
\begin{equation}
D^{-1}\frac{\partial \mu}{\partial t}=L\mu+{\bf R}.
\end{equation}
For the purposes of the current study, L is the Laplacian operator but in future it may be generalised to any linear differential operator on vector-valued functions of vector x.
Now suppose that (\ref{coupled}) allows the reduction
\begin{eqnarray}
\mu_j={(e^{At})_j}^kF_k({\bf x});\\
\label{Helmholtz}
\nabla^2F_k({\bf x})+{M_k}^jF_j({\bf x})=0,
\end{eqnarray}
where $A$ and $M$ are constant matrices. Following that reduction,
\begin{eqnarray}
\nonumber
D^{-1}Ae^{At}{\bf F}&=&Le^{At}\bf{F}+{\bf R}\\
\nonumber
&=&e^{At}L{\bf F}+{\bf R}\\
&=&-e^{At}M{\bf F}+{\bf R}.
\label{twined}
\end{eqnarray}
From here, we need to also assume the commutation property $[A,M]=0$, after which (\ref{twined}) reduces to a constraint among the modelling functions $D(\mu)$ and $R(\mu)$,
\begin{equation}
\label{relnvector}
D^{-1}A{\mu}=-M{\mu}+{{\bf R}}.
\end{equation}
Given that constraint, the system of reaction-diffusion equations is compatible with $\mu_t=A \mu$,
 which may be regarded as the invariant surface condition of a nonclassical symmetry
generated by
\begin{equation}
\label{symmetry}
 \Gamma=\frac{\partial}{\partial t}  +A_1^j\mu_j\frac{\partial}{\partial\mu_1}+A_2^j\mu_j\frac{\partial}{\partial\mu_2} \equiv \frac{\partial}{\partial t}  +(A\mu)\frac{\partial}{\partial\mu},
\end{equation}
where $ \mu=(\mu_1, \mu_2)$ and
 $ \frac{\partial}{\partial\mu}=\Big(\frac{\partial}{\partial\mu_1},\frac{\partial}{\partial\mu_2}\Big)$.
The second prolongation of $\Gamma$ leaves invariant the system of PDEs
 consisting of (\ref{mueq}) together with the vector invariant surface condition.
 So, operator (\ref{symmetry}) is the nonclssical ($Q$-conditional) symmetry. On the other hand,  this operator does not satisfy the classical Lie criteria to be a Lie symmetry. It can happen only for systems of the form  (\ref{mueq}) in exceptional cases. For example, assuming that the matrix $D$ is diogonal, all such systems can be idetified from paper
\cite{ch-king-2006} (see cases 3 and 6 in Table 1 therein).

 In general, sets of nonclassical symmetries do not form a Lie algebra and they cannot be integrated to a Lie group. However in this case of a one-parameter symmetry, invariant solutions are of the form $\mu=e^{At}{\bf F}(x)$ and they are certainly invariant under
$$\bar\mu=e^{\epsilon A}\mu=\mu+\epsilon A\mu+O(\epsilon^2);~~\bar t=t+\epsilon;~~\bar x^i=x^i.$$
{Of course that transformation has no nontrivial action unless it acts on the wider class
 of non-invariant solutions.}\\

Solutions for the flux potentials $\mu_p(x,t)$ can be obtained by solving the linear
Helmholtz system (\ref{Helmholtz}). Solutions $\theta_q(x,t)$ of the reaction-diffusion
 system can be obtained from the flux potentials  provided the Jacobian matrix $\partial \mu_p/\partial \theta_q$ is invertible. That Jacobian is simply the diffusivity matrix $D_p~^q$.\\

 Given the {flux potential functions $\mu_p(\theta)$ and the consequent} diffusivity functions ${D_j}^k(\theta)$,  the partnering reaction terms $R_j({\bf\theta})$ can be determined explicitly from the constraint.  On the other hand if the two reaction terms are specified, then (\ref{relnvector}) is a {system of two first-order partial differential  equations for the partnering potentials $\mu_p(\theta_k)$ that in general will be difficult to solve exactly. Even in the scalar case, the ordinary differential equation for partnering $D$ from $R$ is a difficult Abel equation.}\\

We first consider $A_1^{~2}=-A_2^{~1}=1$ and $A_1^{~1}=A_2^{~2}=0$ as that matrix A would generate interesting oscillations in time as it has eigenvalues $\pm i$. For example, phyto-plankton and zoo-plankton populations have been observed to oscillate \cite{Huisman}. \\

A second case of interest would be a diagonal matrix $A$ with negative eigenvalues. This might represent an ecosystem
 susceptible to species extinction.
In this reduction method, one must find the commutant of $A$, ie the set of all matrices $M$ such that $MA-AM=0$. Then construct the most general form of allowable reaction vectors
$${\bf R}=D^{-1}A\mu+M\mu.$$
Notably, there are many practical applications to heat and mass
transport when $D$ has positive eigenvalues (e.g. \cite{Fulford}).

\section{Oscillatory predator-prey dynamics with spatial structure.}

The simplest way to satisfy (\ref{exact}) is to restrict ${D_j} ^k$ to depend on $\theta_k$ only.
Then
\begin{equation}
\mu_p=\sum_q\int _{\theta_{q0}}^{\theta_q} {D_p}^q(\bar\theta_q) d\bar\theta_q.
\end{equation}

{When considering 2$\times$2 matrices, it is most convenient to use a real basis of Pauli spin matrices including
$$I=\left(\begin{array}{cc}
1 & 0\\
0 & 1
\end{array}
\right),
~\sigma_1=\left(\begin{array}{cc}
0 & 1\\
1 & 0
\end{array}
\right),~i\sigma_2=\left(\begin{array}{cc}
0 & 1\\
-1 & 0
\end{array}
\right),~\sigma_3=\left(\begin{array}{cc}
1 & 0\\
0 & -1
\end{array}
\right).$$
Using that basis it can easily be shown that for any non-singular matrix $A\ne m_0I$, every member $M$ of the commutant of $A$ must be of the form
\begin{equation}
M=m_0I+bA
\end{equation}
with $m_0,b\in \mathbb R.$
Skew symmetric A will have pure imaginary eigenvalues. This will lead to sinusoidal oscillations among the Kirchhoff variables $\mu_j$. In this section it will be assumed that $A=i\sigma_2$ which is a square root of $-I$. Hence
$e^{At}=\cos (t)I +\sin(t)A$ so that
\begin{eqnarray}
\nonumber
\mu_1(x,t)=F_1(x)\cos t+F_2(x)\sin t,\\
\label{rotation}
\mu_2(x,t)=-F_1(x)\sin t+F_2(x)\cos t.
\end{eqnarray}
 Although solutions $\mu_j$ oscillate through positive and negative values, population densities $\theta_i$ cannot take negative values. Therefore the fixed point at $\mu_i=0$ must correspond to  positive-valued  populations  $\theta_i=k_i$ (hereafter the index $i=1,2$).  }\\

Individuals of intelligent species do not move aimlessly but they respond to locations of other species in their food chain. Consider a predator-prey system in which the flux densities of predators and prey are respectively
\begin{equation}
{\bf J}^1=-\nabla\mu_1;~~\mu_1=d_{12}(\theta_2^{\lambda_2}-k_2^{\lambda_2})/\lambda_2
\end{equation}
and
\begin{equation}{\bf J^2}=-\nabla\mu_2;~~\mu_2=d_{21}(\theta_1^{\lambda_1}-k_1^{\lambda_1})/\lambda_1
\end{equation}
with $\lambda_j>0,~k_j>0,~ d_{12}<0$ and $d_{21}>0$. This means that predators will migrate towards higher densities of prey whereas prey will migrate away from higher densities of predators.

This leads to a power-law cross-diffusion matrix
\begin{equation}
\mathcal D=\left(\begin{array}{cc}
0 & d_{12}\theta_2^{\lambda_2-1}\\
d_{21}\theta_1^{\lambda_1-1} & 0
\end{array}
\right).
\end{equation}
Choose $M=0$ and $A=i\sigma_2$.
The consistency relations (\ref{relnvector}) for the nonclassical reduction {require
\begin{equation}
\left(\begin{array}{c}
R_1\\
R_2\end{array}\right)=
\frac{1}{d_{12}d_{21}}\left(\begin{array}{cc}
0 & d_{12}\theta_1^{1-\lambda_1}\\
d_{21}\textcolor{red}{\theta_2}^{1-\lambda_2} & 0
\end{array}
\right)\left(\begin{array}{c}
\mu_2\\
-\mu_1
\end{array}\right).
\end{equation}
In order to have non-singular reaction terms $R_j$ that depend on both populations, $\lambda_j\in (0,1).$  A particular amenable model occurs when $\lambda_1=\lambda_2=\frac 12$, leading to the system
\begin{eqnarray}
\nonumber
\frac{\partial \theta_1}{\partial t}=-2|d_{12}|\nabla\cdot\nabla \theta_2^{1/2}-2\frac{|d_{12}|}{|d_{21}|}k_2^{1/2}\theta_1^{1/2}+
2\frac{|d_{12}|}{|d_{21}|}\theta_1^{1/2}\theta_2^{1/2},\\
\label{3-4}
\frac{\partial \theta_2}{\partial t}=2|d_{21}|\nabla\cdot\nabla \theta_1^{1/2}+2\frac{|d_{21}|}{|d_{12}|}k_1^{1/2}\theta_2^{1/2}-2\frac{|d_{21}|}{|d_{12}|}
\theta_2^{1/2}\theta_1^{1/2}.
\end{eqnarray}
The reaction terms here are comparable to those of the standard
Lotka-Volterra predator-prey system which has
$R_1=-p_1\theta_1+s_1\theta_1\theta_2$ for the predator and
$R_2=p_2\theta_2-s_2\theta_1\theta_2$ for the prey. After the
transformation $\phi_i=\sqrt \theta_i$, the steady states for
$\phi_i(x)$ are exactly the same as those of the standard diffusive
Lotka-Volterra system. While the stability status of those steady
states will be the same, the growth and decay rates of perturbations
will be significantly different. For the standard Lotka-Volterra
model, in the absence of predators, the prey population has
unrestricted exponential growth due to a constant logarithmic growth
rate $\frac{\partial\theta_2}{\partial t}/\theta_2=p_2$. Although in
the current power-law model, in the absence of predators the prey
population
 still has no bounding carrying capacity, growth in this case is more realistic as the logarithmic
 growth rate approaches zero as $\theta_2$ increases:
\begin{equation}
\frac{1}{\theta_2}\frac{\partial\theta_2}{\partial t}={2}\frac{|d_{21}|}{|d_{12}|}k_1^{1/2}\theta_2^{-1/2}.
\end{equation}
\\
   With $M=0$, $F_j(x)$ can be any harmonic functions (see (\ref{Helmholtz})).
Having the  correctly-specified harmonic  functions,
the functions $\mu_j( x,t)$ are then given explicitly by (\ref{rotation}).
Therefore the functions  $\theta_j$  can be found as the   explicit functions of $(x,t)$:
\begin{eqnarray}
\nonumber
\theta_1(x,t)=\left(\frac{\mu_2}{2d_{21}}+\sqrt k_1 \right)^2,\\
\label{3-6}
\theta_2(x,t)=\left(\frac{\mu_1}{2d_{12}}+\sqrt k_2 \right)^2.
\end{eqnarray}

{
Now we present an example, which describes the prey-predator interaction  based on the nonlinear model (\ref{3-4}).
Obviously, $(k_1, k_2)$ is a steady state point of (\ref{3-4}). It can be checked that it is a center similarly to the case of the standard  predator-prey system . We specify this point as $(1, 1)$ in what follows  and set $-d_{12}=d_{21}=1/2$ (just to simplify the calculations) , i.e. examine the system
\begin{eqnarray}
\nonumber
\frac{\partial \theta_1}{\partial t}=-\nabla^2 \theta_2^{1/2}-2\theta_1^{1/2}+
2\theta_1^{1/2}\theta_2^{1/2},\\
\label{3-7}
\frac{\partial \theta_2}{\partial t}=\nabla^2 \theta_1^{1/2}+2\theta_2^{1/2}-2
\theta_2^{1/2}\theta_1^{1/2}.
\end{eqnarray}
Let us specify also the domain, in which two populations interact   as $\Omega=\left\{ (t,x_1,x_2) \in [0,+ \infty )\times(0, \pi)^2\right\}$. Assuming the zero flux conditions on the boundaries, excepting the piece $x_2=0$, where the densities of both  populations can be artificially regulated as periodic functions in time,  we arrive at the boundary conditions
\begin{eqnarray}
\nonumber
&x_1&=0:  \frac{\partial \theta_1}{\partial x_1}=0,  ~~\frac{\partial \theta_2}{\partial x_1}=0,  \\
&x_1&=\pi :  \frac{\partial \theta_1}{\partial x_1}=0, ~~ \frac{\partial \theta_2}{\partial x_1}=0, \\
 \nonumber
&x_2&=0: \, \frac{\partial \theta_1}{\partial x_2}=0, ~~ \frac{\partial \theta_2}{\partial x_2}=0, \\
\nonumber
&x_2&=\pi: \,  \theta_1 = (-f_1\sin t+f_2\cos t  + 1)^2,\\
\label{3-8}
  &~~& ~~~~~~~~  \theta_2 = (f_1\cos t  +f_2\sin t+ 1)^2
\end{eqnarray}
where $f_1(x_1)$  and $f_2(x_1)$ are given functions.
\begin{figure}\begin{center}
 \includegraphics[width=7cm]{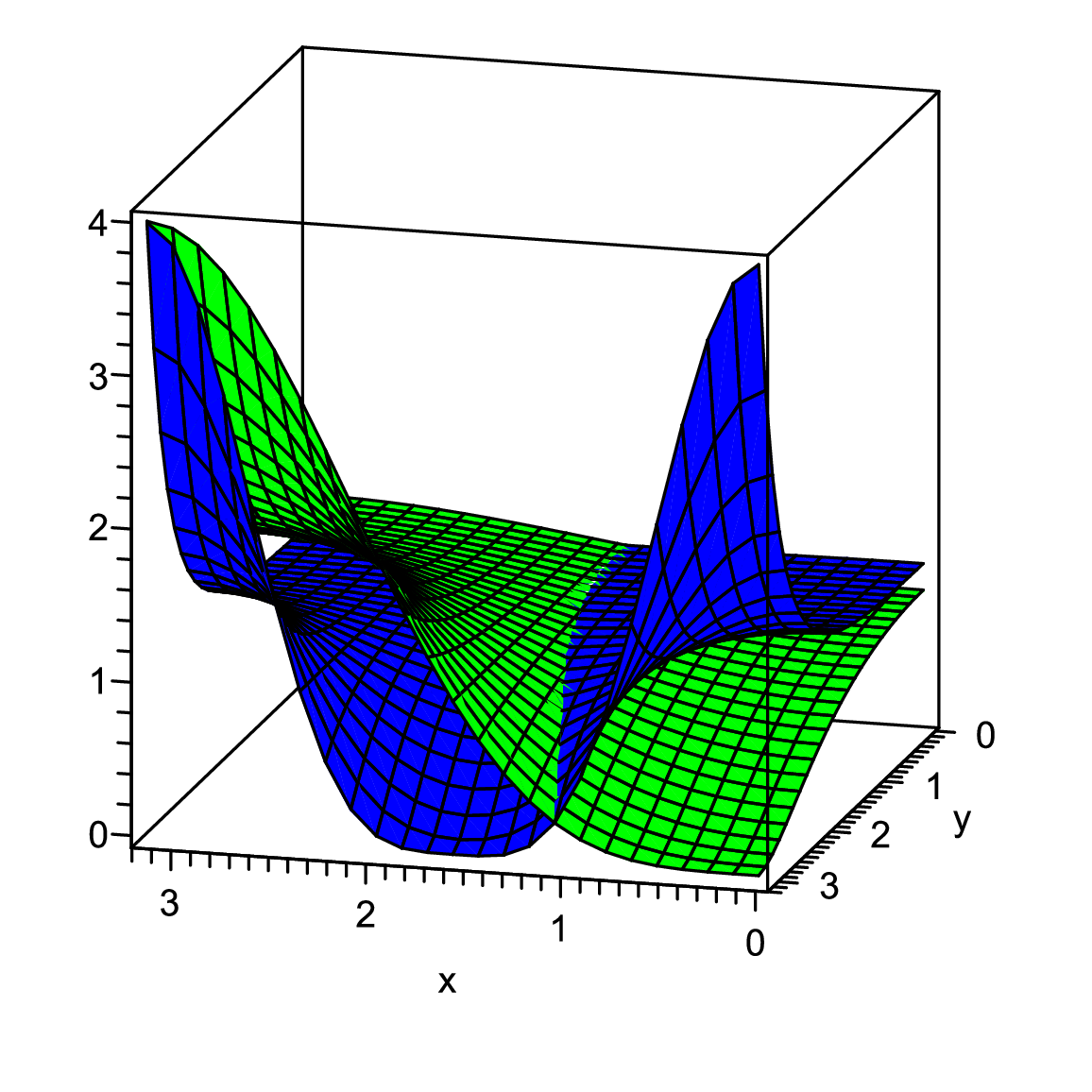}
 \end{center}
 \caption{$t=0$,  $x=x_1, \  y=x_2$}\label{f1}
\end{figure}

\begin{figure}\begin{center}
 \includegraphics[width=7cm]{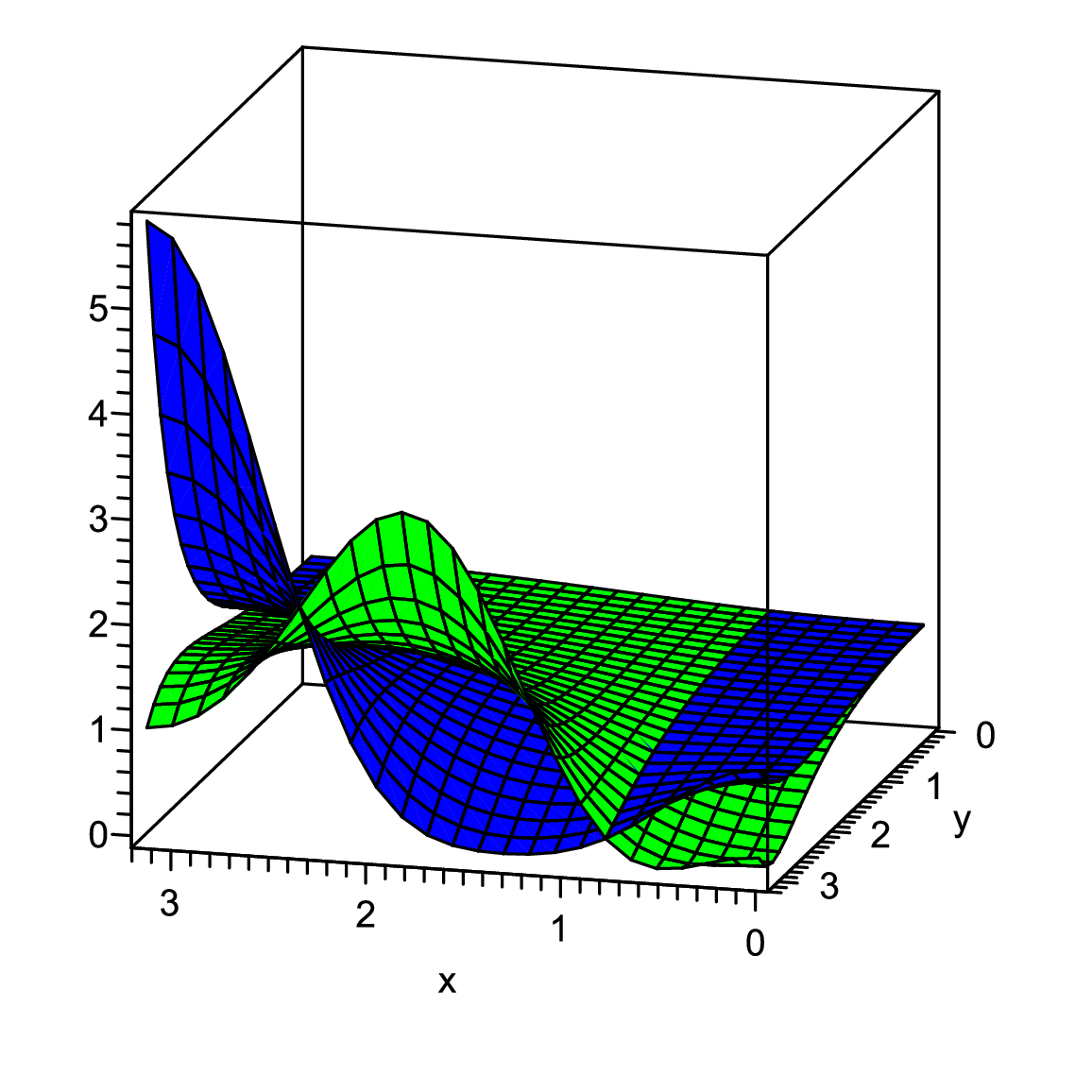}
 \end{center}
 \caption{$t=\frac{\pi}{4},$   $x=x_1, \  y=x_2$}\label{f2}
\end{figure}

\begin{figure}\begin{center}
 \includegraphics[width=7cm]{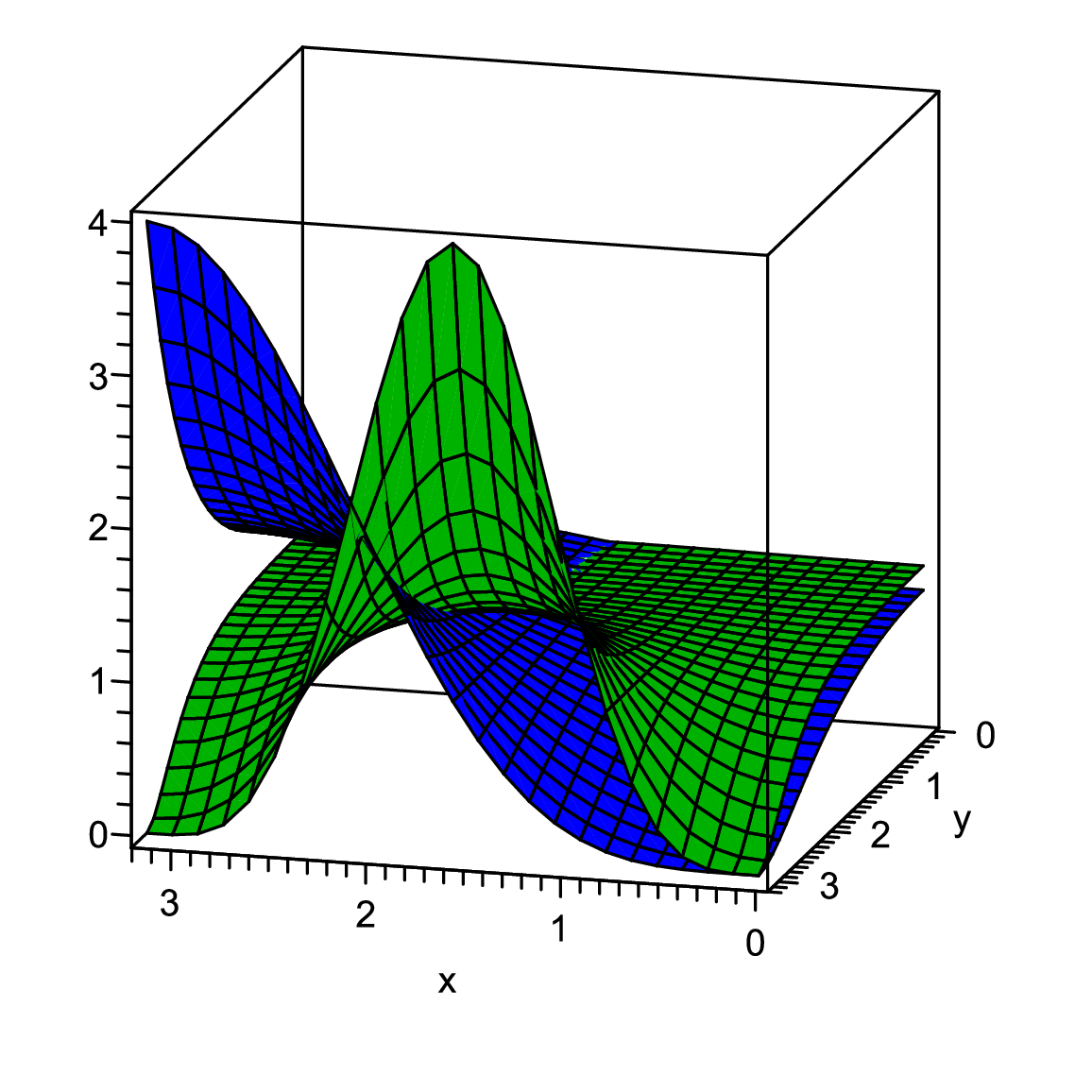}
 \end{center}
 \caption{$t=\frac{\pi}{2}$,   $x=x_1, \  y=x_2$}\label{f3}
\end{figure}

\begin{figure}\begin{center}
 \includegraphics[width=7cm]{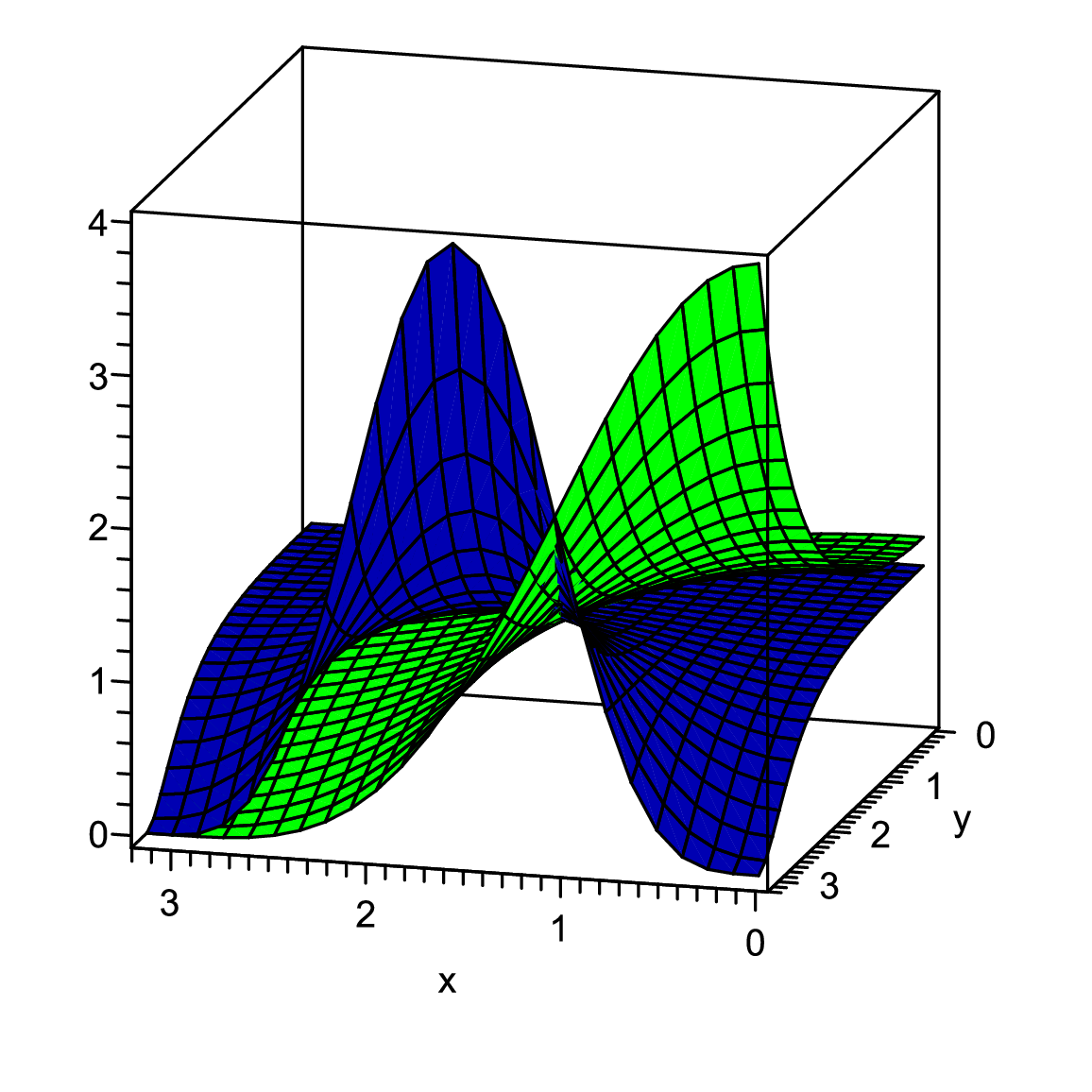}
 \end{center}
 \caption{$t=\pi$,  $x=x_1, \  y=x_2$}\label{f4}
\end{figure}
In order to construct the exact solution of (\ref{3-7}) that satisfies the boundary conditions (\ref{3-8}), we need to solve the linear  boundary value problem
\begin{eqnarray}\label{3-9}
\nabla^2 F_1=0,  \,  \nabla^2 F_2=0,
\end{eqnarray}
and
\begin{eqnarray}
\nonumber
x_1=0: \, \frac{\partial F_1}{\partial x_1}=0, \, \frac{\partial F_2}{\partial x_1}=0,  \,
x_1=\pi: \, \frac{\partial F_1}{\partial x_1}=0, \, \frac{\partial F_2}{\partial x_1}=0,  \\
\label{3-10}
x_2=0: \, \frac{\partial F_1}{\partial x_2}=0, \, \frac{\partial F_2}{\partial x_2}=0,  \,
x_2=\pi: \,  F_1 = f_1(x_1),  \,   F_2 = f_2(x_1).
\end{eqnarray}
It can be done using the classical Fourier method.  As a result, we arrive at the exact solution
\begin{eqnarray}\label{3-11}
F_1= \sum^\infty_{n=0} \frac{a_{1n}}{\cosh (n\pi)} \cos(nx_1)\cosh (nx_2),  \,  F_2= \sum^\infty_{n=0}  \frac{a_{2n}}{\cosh (n\pi)} \cos(nx_1)\cosh (nx_2),
\end{eqnarray}
where $a_{1n}$  and $a_{2n}$  are the Fourier coefficients for the functions $f_1$ and $f_2$, respectively. Thus, the explicit expressions for the functions $\theta_1$  and $\theta_2$ are derived,  using  formulae (\ref{rotation}), (\ref{3-6})  and (\ref{3-11}).

There are  interesting  cases when the infinite series degenerate to just one or two term(s) in (\ref{3-11}).
We could set, for example, $f_j=\cos(jx_1), j=1,2$, therefore the expressions in (\ref{3-11})  take the form
\begin{eqnarray}\label{3-12}   F_1= \frac{\cosh(x_2)}{\cosh(\pi)} \cos(x_1),  \,
 F_2= \frac{\cosh (2x_2)}{\cosh(2\pi)} \cos(2x_1). \end{eqnarray}
Thus, inserting the functions $F_1$ and $F_2$  into (\ref{rotation})  and using the substitution (\ref{3-6}), we arrive at the exact solution of the boundary-value problem (\ref{3-7})--(\ref{3-8})
\begin{eqnarray}
\nonumber
\theta_1 = \Big(1 - \frac{\cosh(x_2)}{\cosh(\pi)} \cos(x_1)\sin t+ \frac{\cosh (2x_2)}{\cosh(2\pi)} \cos(2x_1)\cos t  \Big)^2 \\
\label{3-14}
\theta_2 = \Big(1 - \frac{\cosh(x_2)}{\cosh(\pi)} \cos(x_1)\cos t- \frac{\cosh (2x_2)}{\cosh(2\pi)} \cos(2x_1)\sin t  \Big)^2
 \end{eqnarray}
 We present the  densities  $\theta_1$  and $\theta_2$ in Fig.1, 2, 3 and 4 for the  time moments $t=0, \  t=\frac{\pi}{4}, \ t=\frac{\pi}{2}$  and $t=\pi$, respectively. The blue surface represents  the predator density $\theta_1$, while the green one is for the prey density $\theta_2$.
\medskip\\
In the Lagrangian formulation of mass transport equations, the
trajectories ${\bf r}(t)$ of material particles following the flow,
satisfy the system of differential equations
 $$\frac{d{\bf r}}{dt}=\frac{{\bf j}({\bf r},t)}{\theta({\bf r},t)}$$ where ${\bf j}$ is the mass flux density and $\theta$ is the mass
  density. In the current example, since ${\bf j}$ depends explicitly on time, the flow lines that are integral curves will not be
  the same as the streamlines at constant $t$.
\medskip\\
In this continuum model, the flow lines of small compact assemblies
of fish follow a system of non-autonomous nonlinear differential
equations.
\begin{eqnarray}
\hbox{Species 1:}~~\frac{dx}{dt}=\frac{-\nabla\mu_1}{(1+\mu_2)^2},\\
\hbox{Species 2:}~~\frac{dx}{dt}=\frac{-\nabla\mu_2}{(1-\mu_1)^2}.
\end{eqnarray}
In order to illustrate the oscillatory nature of the flow lines, with ${\bf r}(t)=(x(t),y(t))$ we consider the example
\begin{eqnarray}
\mu_1=\alpha e^{-x}\sin (y+t),\\
\mu_2=\alpha e^{-x}\cos (y+t),
\end{eqnarray}
with $\alpha=0.35$. A time integrated flow line for Species 2, along
with some contours for Species 1 density at a particular time, are
shown in Figure \ref{flow2}. At each time, the flux vector for
predator/prey is normal to the contour line for prey/predator.
However those contours are changing in time, resulting in temporal
oscillations  in the directions of flux vectors.} An integrated flow
path for Species 1 is shown in Figure \ref{flow}.

\begin{figure}\begin{center}
\includegraphics[width=8cm]{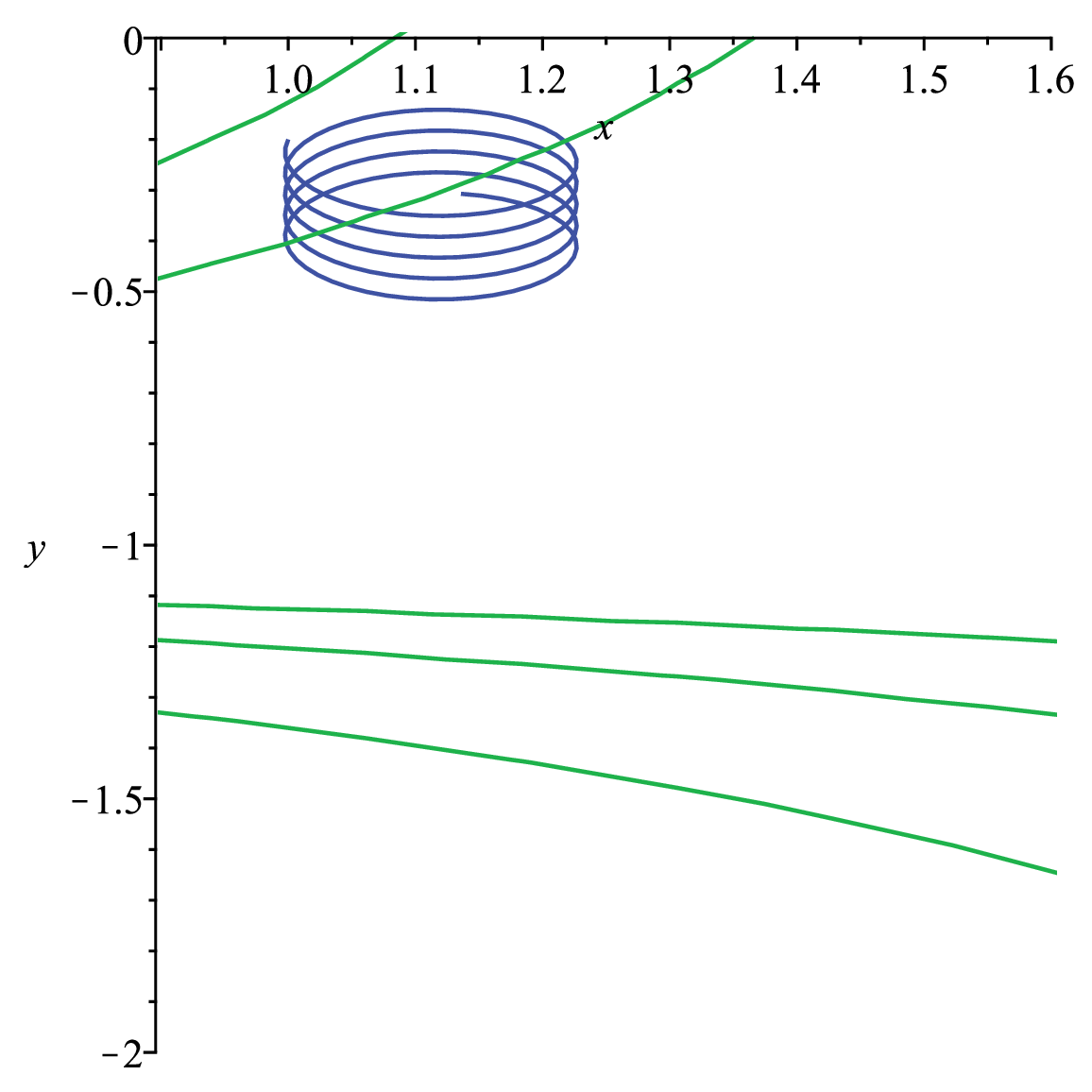}
 \end{center}
 \caption{At $t=7\pi/3$, contours of prey population $\theta_2$=0.9, 0.924, 1.01, 1.02 ,1.04 with flow line of predator Species 1 from t=0 to 30, initial value $(x,y)=(1,-0.2)$.}
\label{flow2}
\end{figure}

\begin{figure}\begin{center}
\includegraphics[width=8cm]{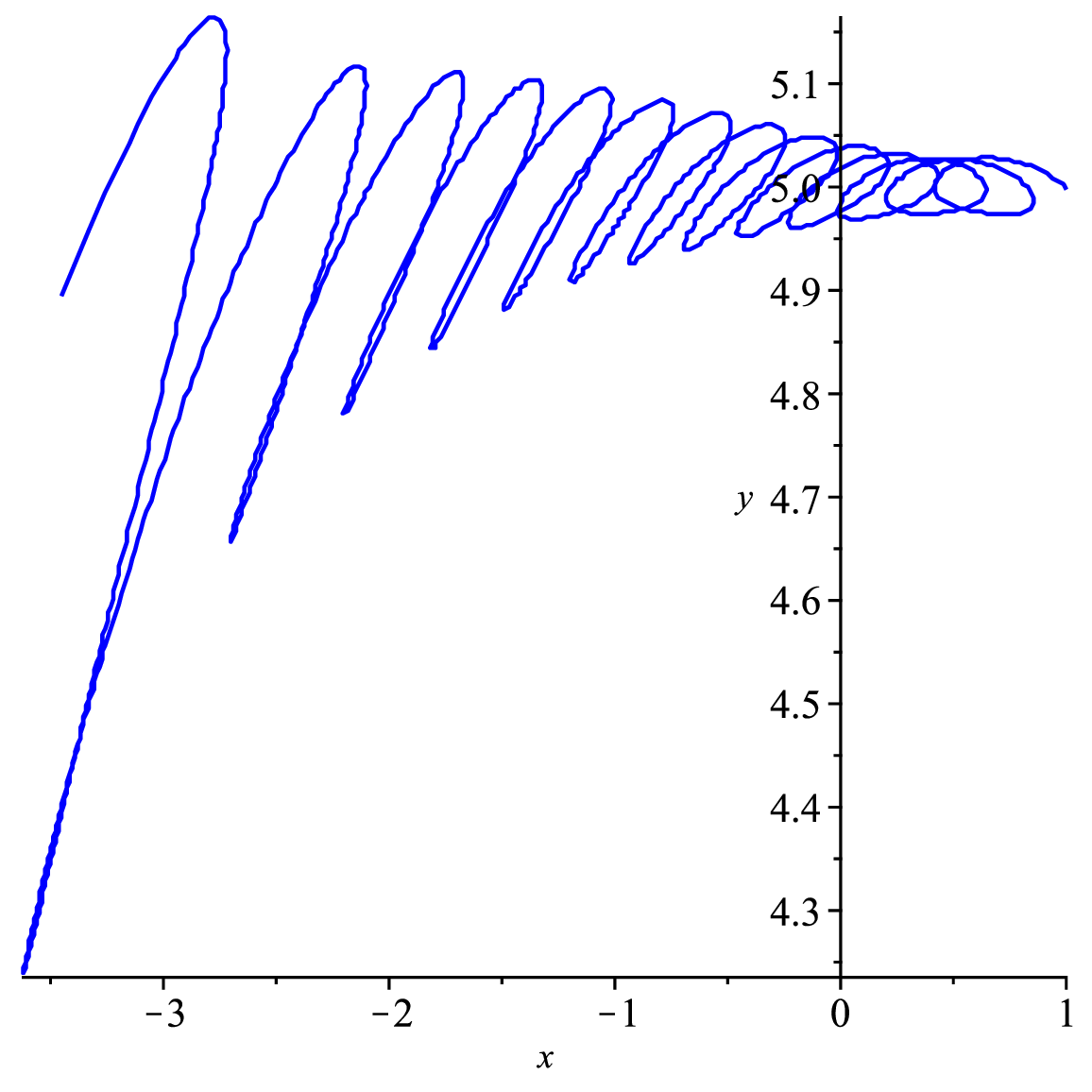}
 \end{center}
 \caption{Flow path for predator Species 1 as in Eq. (37) with $\mu_1=-2\sqrt{\theta_2}+2$, $\mu_2=2\sqrt{\theta_1}-2$ with initial values $(x,y)=(1,5)$.}
\label{flow}
\end{figure}

Finally, we construct a simple example in which $F_1$ and $F_2$ are
the standard velocity potentials of point vortices at locations
$(0,0)$ and $(1,0)$. Although the separation between the vortices is
constant, in the construction of reaction-diffusion solutions, the
flux vector of each species oscillates between that controlled by
$F_1$ and that controlled by $F_2$. Then we take \begin{eqnarray}
F_1(x,y)=\frac 1\pi\tan ^{-1}\frac xy,\\
F_2(x,y)=\frac 1\pi\tan ^{-1}\frac{x-1}{y}.
\end{eqnarray}
Then Species 1 (predator) flow lines are the integral curves of the ODE system
\begin{eqnarray}
\nonumber
\frac{dx}{dt}=\pi\left[\frac{-y\cos t}{x^2+y^2}-\frac{y\sin t}{(x-1)^2+y^2}\right]/\left[\cos (t)\tan ^{-1}\frac{x-1}{y}-\sin (t) \tan ^{-1}\frac xy +\pi\right]^2,\\
\label{PredDE}
\frac{dy}{dt}=\pi\left[\frac{x\cos t}{x^2+y^2}+\frac{(x-1)\sin t}{(x-1)^2+y^2}\right]/\left[\cos (t)\tan ^{-1}\frac{x-1}{y}-\sin (t)\tan ^{-1}\frac xy +\pi\right]^2.
\end{eqnarray}
Similarly the Species 2 (prey) flow lines satisfy
\begin{eqnarray}
\nonumber
\frac{dx}{dt}=\pi\left[\frac{-y\cos t}{(x-1)^2+y^2}+\frac{y\sin t}{x^2+y^2}\right]/\left[\sin (t)\tan ^{-1}\frac{x-1}{y}+\cos (t) \tan ^{-1}\frac xy -\pi\right]^2,\\
\label{PreyDE}
\frac{dy}{dt}=\pi\left[\frac{-x\sin t}{x^2+y^2}+\frac{(x-1)\cos t}{(x-1)^2+y^2}\right]/\left[\sin (t)\tan ^{-1}\frac{x-1}{y}+\cos (t) \tan ^{-1}\frac xy -\pi\right]^2.
\end{eqnarray}
An example of a flow line for each species is given in Figure (\ref{PredPreyFlow}). They were obtained by the ODE solver ODE45 of MATLAB2020b.
\begin{figure}\begin{center}
\includegraphics[width=8cm]{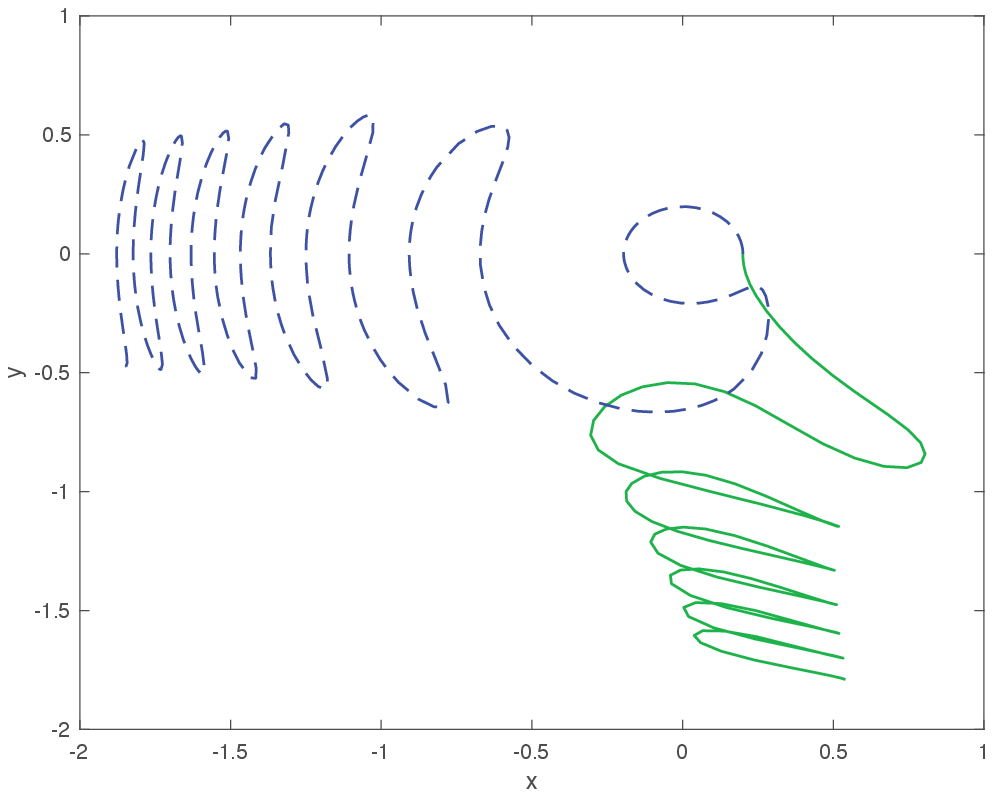}
 \end{center}
 \caption{Flow line (dashed) for predator Species 1 and for prey Species 2 (filled) following ODE systems (\ref{PredDE}) and (\ref{PreyDE}), both with initial values $(x,y)=(0.2,0)$.}
\label{PredPreyFlow}
\end{figure}

\section{Solutions with approach to extinction}
The previous section began by considering  additively separable flux
potential functions $\mu_i(\theta_1,\theta_2)$. Meaningful models
can also be based on multiplicatively separable potential functions
\begin{equation}
\mu_i(\theta_1,\theta_2)=H_i(\theta_1)G_i(\theta_2)~;~~i=1,2.
\end{equation}
In the following example we choose
\begin{eqnarray}
\nonumber
H_1(\theta_1)~~~~=\theta_1^{1/2}~;~~&G_1(\theta_2)&=k_2^{1/2}-\theta_2^{1/2},\\
H_2(\theta_1)=\theta_1^{1/2}-k_1^{1/2};~~&G_2(\theta_2)&=\theta_2^{1/2},
\end{eqnarray}
as well as diagonal matrices
\begin{eqnarray}A=\left(\begin{array}{cc}
a_1 & 0\\
0 & a_2
\end{array}
\right),\\
~M=\left(\begin{array}{cc}
m_1 & 0\\
0 & m_2
\end{array}
\right).
\end{eqnarray}
The diffusion matrix is
\begin{equation}\label{4-1}
\mathcal D= \frac{1}{2}\left(\begin{array}{cc}
\theta_1^{-1/2}(k_2^{1/2}-\theta_2^{1/2}) & -\theta_1^{1/2}\theta_2^{-1/2}\\
\theta_1^{-1/2}\theta_2^{1/2} & \theta_2^{-1/2}(\theta_1^{1/2}-k_1^{1/2})
\end{array}
\right).
\end{equation}

The flux densities of the two species are
\begin{eqnarray}
j_1=-\nabla\mu_1=(\theta_2^{1/2}-k_2^{1/2})\nabla\theta_1^{1/2}+\theta_1^{1/2}\nabla \theta_2^{1/2},\\
j_2=-\nabla\mu_2=(k_1^{1/2}-\theta_1^{1/2})\nabla\theta_2^{1/2}-\theta_2^{1/2}\nabla \theta_1^{1/2}.
\end{eqnarray}
The cross diffusion terms indicate that Species 1 is the predator and Species 2 is the prey. Species 2 avoids regions of relatively high densities of Species 1 whereas Species 1 is attracted towards regions of relatively higher density of Species 1. In the neighbourhood of $\theta=(0,0)$, the self-diffusion coefficients are positive for the predator and negative for the prey. This may correspond to herding or schooling of prey when predators are scarce.\\

Now from the constraint (\ref{relnvector}), reduction to the vector linear Helmholtz equation will be possible when
\begin{eqnarray}
\label{R1mod}
R_1=m_1\theta_1^{1/2}(k_2^{1/2}-\theta_2^{1/2})+2\theta_1(\theta_1^{1/2}-k_1^{1/2})
\frac{\textcolor{blue}{a_1(k_2^{1/2}-\theta_2^{1/2})+a_2\theta_2^{1/2}}}
{(k_2\theta_1)^{1/2}+(k_1\theta_2)^{1/2}-(k_1k_2)^{1/2}},\\
R_2=m_2\theta_2^{1/2}(\theta_1^{1/2}-k_1^{1/2})-2\theta_2(k_2^{1/2}-\theta_2^{1/2})
\frac{\textcolor{red}{a_2(k_1^{1/2}-\theta_1^{1/2})+a_1\theta_1^{1/2}}}
{(k_2\theta_1)^{1/2}+(k_1\theta_2)^{1/2}-(k_1k_2)^{1/2}}.
\label{R2mod}
\end{eqnarray}

Now we choose $m_1<0$ and $m_2<0$ so that these reaction functions have several fundamental features in common with those of
the Lotka-Volterra system and with those of the pursuit model of the previous section:\\
\noindent
(i) the extinction point $\theta=(0,0)$ is a uniform fixed point of the system,\\
(ii) there is exactly one  interior uniform fixed point, namely $\theta=(k_1,k_2)$,\\
(iii) a low  prey population has positive (negative) growth when the
predator population is below (above) some critical
value $k_1$,\\
(iv) a low   predator population has positive (negative) growth when
the prey population is above (below) some critical
 value $k_2$.\\

This model has the advantage of having an additional fixed point on
the zero-predator boundary. In the absence of any
 predators, the prey population has logistic production rate
$$R_2=\frac{|m_2}{2|a_2|}| \phi_1\left[1-\frac{\phi_1}{\theta_c^{0.5}}\right],$$
where $\phi_2=\sqrt\theta_2$ and the carrying capacity is
$\theta_c=m_2^2 k_1/4a_2^2.$ This is an improvement over the
standard diffusive Lotka-Volterra system for which the prey has
unbounded growth with unlimited carrying capacity in the absence of
predators.

Now in the neighbourhood of fixed point ${\bf\theta}={\bf 0}$, at leading order in $ { \theta_j}$,
\begin{eqnarray}
R_1\approx m_1k_2^{1/2}\theta_1^{1/2}<0,\\
R_2\approx -m_2 k_1^{1/2}\theta_2^{1/2}>0.
\end{eqnarray}
This implies that with uniform populations, the zero fixed point will be a saddle point. The predator approaches extinction due to lack of prey but the prey have a positive net {growth rate}.\\
In the neighbourhood of fixed point $(\theta_1,\theta_2)=(k_1,k_2)$, to leading order in $ \theta_j^{1/2}-{ k_j}^{1/2}$,
\begin{eqnarray}
R_1\approx |m_1|k_1^{1/2}(\theta_2^{1/2}-k_2^{1/2})+2a_2k_1^{1/2}(\theta_1^{1/2}-k_1^{1/2})\\
R_2\approx -|m_2| k_2^{1/2}(\theta_1^{1/2}-k_1^{1/2})+2a_1k_2^{1/2}(\theta_2^{1/2}-k_2^{1/2}).
\end{eqnarray}
If $a_1$ and $a_2$ were zero, the leading terms in $R_j$ would lead to this fixed point being a focus for uniform population dynamics, just as in the original Lotka-Volterra system. More generally we allow $a_1$ and $a_2$ to be negative, resulting in the fixed point being a stable focus for uniform population \textcolor{red}{dynamics, with some} inward spiralling orbits due to the cyclic behaviour governed by the $m_j$ terms. We avoid positive values of $a_j$ that would lead to unbounded dynamics in the region $\theta_j>k_j$. When one considers the dependence of populations on both space and time, small perturbations about the fixed point will satisfy a system of linear partial differential equations, not just a system of linear ordinary differential equations. We can actually construct some exact solutions of the full nonlinear system that approach the fixed point. The system $(\ref{Helmholtz})$ consists of two independent modified Helmholtz equations for which exact solutions are readily available.\\

 In order to construct some exact solutions  and provide their possible biological interpretation,  we specify some parameters as follows:
 $k_1=k_2=1$ (i.e. population densities are scaled by their equilibrium values) and
 $2a_1=2a_2=a<0$.
 Henceforth, the reaction terms (\ref{R1mod})-(\ref{R2mod})  take the form
\begin{eqnarray}
\label{4-3}
R_1=m_1\theta_1^{1/2}(1-\theta_2^{1/2})+
\frac{a\theta_1(\theta_1^{1/2}-1)}
{\theta_1^{1/2}+\theta_2^{1/2}-1},\\
R_2=m_2\theta_2^{1/2}(\theta_1^{1/2}-1)-
\frac{a \theta_2(1-\theta_2^{1/2}) }
{\theta_1^{1/2}+\theta_2^{1/2}-1}
\end{eqnarray}
The non-zero fixed point is $(1,1)$. Exact solutions of the relevant boundary-value problem are now constructed as in Section 3.} Notably, the diagonal diffusivities vanish at  the steady-state point $(1,1)$ (see (\ref{4-1}).)

\begin{figure}\begin{center}
\includegraphics[width=8cm]{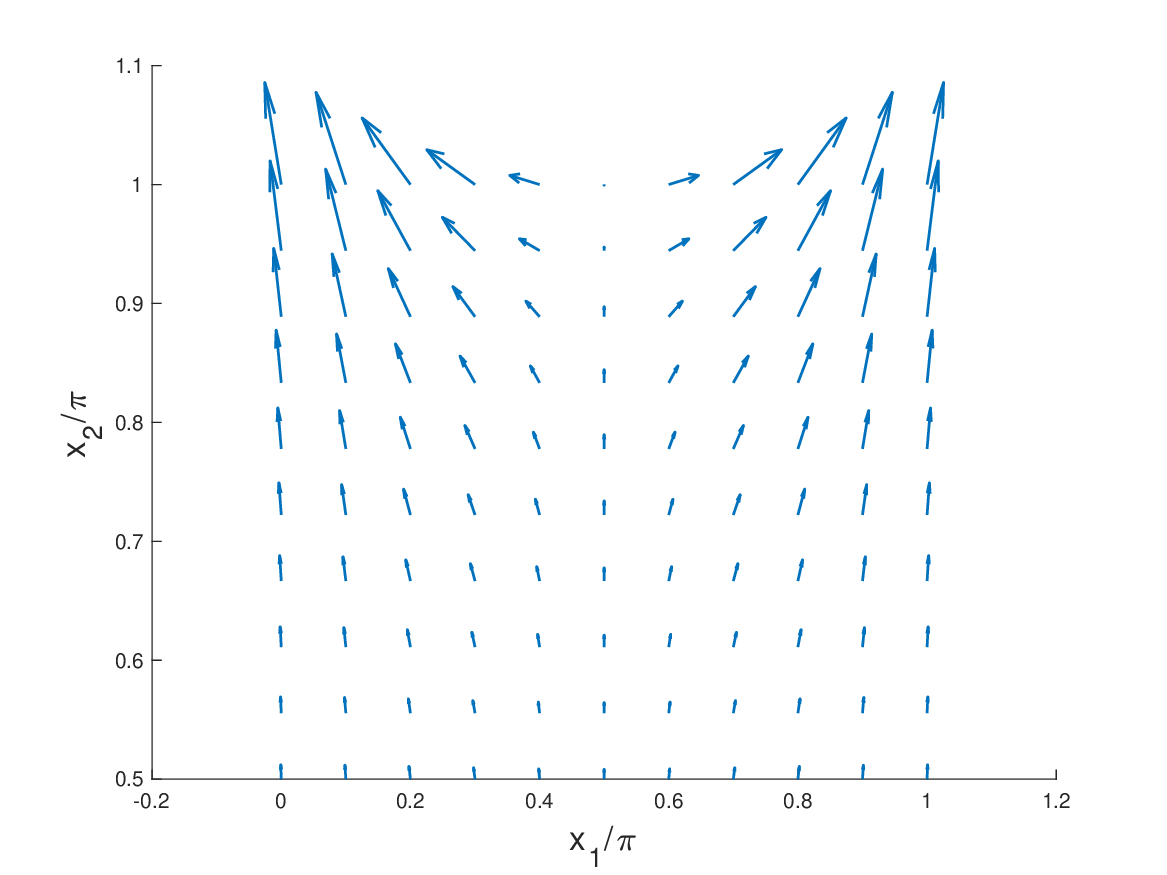}
 \end{center}
 \caption{Initial flux vector field for a predator in a square holding pen with one side open..}
\label{pen}
\end{figure}

 Now assume $k_1=k_2=1$ and $a_1=a_2=2a$.  Three of the sides of a rectangular domain will be assumed to be barriers
  to flow in the normal direction ${\bf n}$ , so that for all $i$, ${\bf n}\cdot\nabla\mu_i =0$, which is equivalent
  to  $\forall i ~{\bf n}\cdot\nabla\theta_i=0.$ The remaining side will have time dependent boundary conditions with populations
  approaching their steady state values: $\forall i ~\mu_i\to 0, ~\theta_i\to1$. These boundary conditions are specified as
{
\begin{eqnarray}
\nonumber
x_1=&0:& \, \frac{\partial \theta_1}{\partial x_1}=0,  ~~\frac{\partial \theta_2}{\partial x_1}=0,  \\
x_1=&\pi : &\, \frac{\partial \theta_1}{\partial x_1}=0, ~~ \frac{\partial \theta_2}{\partial x_1}=0, \\
 \nonumber
x_2=&0:& \, \frac{\partial \theta_1}{\partial x_2}=0, ~~ \frac{\partial \theta_2}{\partial x_2}=0, \\
\nonumber
x_2=&\pi :& \, \theta_1
=\ds \frac{1}{4}\left[ {(1 +e^{2at} (f_1+f_2)) \pm \sqrt{(1 +e^{2at} (f_1+f_2))^2 - 4e^{2at} f_1}} \right ]^2, \\
\nonumber
x_2=&\pi:& \, \theta_2
=\ds \frac{1}{4}\left[ {(1 -e^{2at} (f_1+f_2)) \pm \sqrt{(1 -e^{2at} (f_1+f_2))^2 + 4e^{2at} f_2}} \right ]^2,\\
\label{54}
\end{eqnarray}
where $f_1(x_1)$  and $f_2(x_1)$ are given functions. In the
following, since we are considering solutions in the neighbourhood
of the non-zero interior fixed point, we choose the larger root for
$\theta_1$ (with the + sign alternative).} We have that
 \beq \mu_1= \theta_1^{1/2}(1-\theta_2^{1/2}), \ \ \mu_2=(\theta_1^{1/2}-1)\theta_2^{1/2}, \label{4-5} \eeq
and with $A=2aI$, that $\exp(At)=\exp(2at)I$ so
\beq  \mu_1 = e^{2at} F_1(x_1,x_2), \ \
\mu_2 =  e^{2at} F_2(x_1,x_2), \label{4-6} \eeq
where
\begin{eqnarray}\label{4.7}
\nabla^2 F_1 +m_1F_1=0,  \,  \nabla^2 F_2+m_2F_2=0,\ \ \  m_1<0,m_2 <0.
\end{eqnarray}
The relevant boundary conditions in terms of $F_1$ and $F_2$ are
\begin{eqnarray}
\nonumber
x_1=0&:&  \frac{\partial F_1}{\partial x_1}=0, \, \frac{\partial F_2}{\partial x_1}=0,  \\
x_1=\pi&:&  \frac{\partial F_1}{\partial x_1}=0, \, \frac{\partial F_2}{\partial x_1}=0,  \\
\label{4.8}
x_2=0&:&  \frac{\partial F_1}{\partial x_2}=0, \, \frac{\partial F_2}{\partial x_2}=0,  \\
x_2=\pi &:&   F_1 = f_1(x_1), \ \   F_2 = f_2(x_1).\
 \end{eqnarray}
We solve for $F_1$ and $F_2$ using the classical Fourier method.  As a result, we arrive at the exact solution
\begin{eqnarray}\label{4.9}
F_1=\frac{\alpha_0}{2}+ \sum^\infty_{n=1} \alpha_n \cos nx_1 \ \cosh(\sqrt{n^2-m_1} x_2),
\end{eqnarray}
 where $\alpha_n \cosh(\sqrt{n^2-m_1} \pi) = {2\over \pi} \int_0^\pi \cos nx_1 \ f_1( x_1) \ dx_1.$
Similarly,
 \begin{eqnarray}\label{4.10}
F_2= \frac{\gamma_0}{2}+\sum^\infty_{n=1} \gamma_n \cos nx_1 \ \cosh(\sqrt{n^2-m_2} x_2),
\end{eqnarray}
 where $\gamma_n \cosh(\sqrt{n^2-m_2} \pi) = {2\over \pi} \int_0^\pi \cos nx_1 \ f_2( x_1) \ dx_1$ .
 Equating (\ref{4-5}) and (\ref{4-6}) we have that
 \beq \theta_1-\theta_1^{1/2}(1+e^{2at}F_1+e^{2at} F_2)+e^{2at}F_1=0, \eeq
 so that {
  \beq \theta_1=\frac{1}{4}\left[{(1+e^{2at}F_1+e^{2at} F_2)+ \sqrt{(1+e^{2at}F_1+e^{2at} F_2)^2-4e^{2at}F_1}}\right ]^2.\eeq
Similarly,
  \beq \theta_2=\frac{1}{4}\left[{(1-e^{2at}F_1-e^{2at} F_2)+ \sqrt{(1-e^{2at}F_1-e^{2at} F_2)^2+4e^{2at}F_2}}\right ]^2.\eeq
 \label{thetaF} }
{\textit{Note 1:}} If $f_1(x_1)= f(x_1)$ and $f_2(x_1)=-f(x_1)$ then
$\ds \theta_1(t,x_1,\pi)= \theta_2(t,x_1,\pi)
= \left ({1+ \sqrt{1-4e^{2at}f}\over 2} \right )^2.$\\
{\textit{Note 2:}} In (\ref{54}) $a$ must be negative so the
populations approach their fixed points. In order to build from a
simple example, we begin with initial uniform values at the open
boundary, $\theta_1(x_1,\pi, 0)=\frac 12$ and
$\theta_2(x_1,\pi,0)=\frac 32$. This corresponds to a predator
population initially below its steady value and a prey population
initially above its steady value. From (\ref{thetaF}), this
corresponds to negative boundary values for $F_i$, $f_1(x_1)=(\sqrt
2-\sqrt 3)/2$ and $f_2=(\sqrt 3-\sqrt 6)/2$. From the solution it
can be seen that the populations asymptotically approach their
steady state values $1$ everywhere. Since there is no variation in
the $x_1$ direction, this is so far a one-dimensional problem. It
can be made a two-dimensional problem simply by adding other Fourier
components in $f_i$. We consider the example,
$$f_1(x_1,x_2)=-0.5[\sqrt 3-\sqrt 2][1+0.5\cos(x_1)];~~f_2(x_1,x_2)=\left(1-\frac{\sqrt 6}{2}\right)f_1(x_1,x_2).$$
The initial flux vector of the predator  in the upper half of the square domain is depicted in Figure (\ref{pen}).
In the lower half, fluxes are close to zero.
Individuals of the predators and prey initially escape out of the open end of the square holding pen, towards the corners. The flux approaches zero as the predator density approaches its steady value 1 from below and the prey density approaches its steady value 1 from above.\\
With $m_i<0$, $\mu_i$ satisfies the modified Helmholtz equation.
Exact solutions can be readily constructed by substituting a pure
imaginary valued wave number in many standard solutions of the usual
Helmholtz equation that occurs in acoustic scattering theory (e.g.
\cite{Philip}).

\normalcolor

\section{Conclusion}
Here we have demonstrated a highly unusual circumstance of a conditionally integrable system of
two nonlinear partial differential equations in one time and $N$ space dimensions.
Via a nonclassical symmetry, that \textit{nonlinear} system reduces to a \textit{linear} system of
two coupled Helmholtz equations in $N$ space dimensions. From there we can construct an infinite
dimensional linear space of solutions that depend on both space and time,
 which is a proper sub-manifold within the larger infinite dimensional manifold of solutions of
  the original nonlinear system. \\

For purposes of illustration, this paper has focused on coupled nonlinear reaction-diffusion equations in two space dimensions.
 The technique requires the original nonlinear system to be augmented by one of a 4-parameter set of possible side conditions that
 relate the nonlinear diffusion matrix to the nonlinear source vector.  Exact solutions of diffusive predator-prey systems have been
  constructed, some that decay towards extinction and some that oscillate or spiral around an interior fixed point. The conditionally
  integrable systems are closely related to the standard Lotka-Volterra system but they have two additional features that are advantageous.
   Firstly, unlike in the standard system, in the absence of predators the expanding prey population does not exhibit un-natural unbounded
    exponential growth but it may have a carrying capacity, as in the diffusive Fisher equation. Secondly, unlike the standard predator-prey
     system, the nonclassical reduction method makes available a wide variety of exact solutions that vary in both space and time.
      For example when constructing solutions that are oscillatory in time, a different solution can be constructed from any pair of
      solutions of Laplace's equation, not necessarily conjugate harmonic pairs. We have explicitly calculated fluxes and densities,
      analogous to the so-called Euler picture of fluid mechanics.  From the Euler picture, we have constructed the alternative Lagrange
       picture that is a system of nonlinear non-autonomous ordinary differential equations. Their integral curves, obtained numerically
       here, are sample paths of individual elements of the predator and prey populations, down to the individual or small group level.
       These are analogous to the flow lines in fluid mechanics, as opposed to the stream lines that are vector fields that are frozen at
       a particular time. \\

In the examples of solutions that we have constructed we have not
yet maintained standard boundary conditions on all of the boundary
of the domain. In principle, some examples of standard boundary
value problems might be attained by such methods as conformal
mapping and classical scattering techniques that apply to the Laplace and Helmholtz systems that are obtained by reduction.\\

As we have previously seen in applications to scalar equations, the
target nonlinear PDEs may potentially involve not only reaction and
diffusion terms but also convection terms and higher-order
diffusion. From an imposed nonlinear diffusivity matrix, the
construction of  compatible source terms in a conditionally
integrable model is straightforward. Unlike in the nonclassical
symmetry
 reduction of a scalar PDE, as yet we know of no simple method to construct a partner diffusion matrix from imposed reaction functions.
  That is an important problem whose solution would lead to insight on a wide range of physical applications.\\

\end{document}